\documentclass[showpacs,pre,aps,twocolumn,superscriptaddress]{revtex4-1}

\pdfoutput=1
\usepackage{amsmath,amsfonts,amssymb,bm,graphicx,color}

\newcommand{\C}{\mathcal{C}}
\newcommand{\II}{\mathcal{I}}
\newcommand{\QQ}{\mathcal{Q}}
\newcommand{\WW}{\mathcal{W}}
\newcommand{\RR}{\mathcal{R}}
\newcommand{\TT}{\mathcal{T}}
\newcommand{\W}{\mathbb{W}}
\newcommand{\T}{\mathbb{T}}
\newcommand{\Q}{\mathbb{Q}}
\newcommand{\I}{\mathbb{I}}
\newcommand{\M}{\vec{M}}
\newcommand{\s}{\vec{s}}
\newcommand{\m}{\vec{m}}
\newcommand{\Y}{{\bf Y}}

\newcommand{\llangle}{\langle\!\langle}
\newcommand{\rrangle}{\rangle\!\rangle}
\newcommand{\bra}[1]{\langle #1|}
\newcommand{\ket}[1]{|#1\rangle}

\newcommand{\Tr}{\mathop{\mathrm{Tr}}}

\newcommand{\tobs}{\tau}

\begin{document}

\title{Fluctuating observation time ensembles in the thermodynamics of trajectories}

\author{Adri\'an A. Budini}
\affiliation{Consejo Nacional de Investigaciones Cient\'{\i}ficas y T\'{e}cnicas (CONICET), Centro At\'{o}mico Bariloche, Avenida E. Bustillo Km 9.5, (8400) Bariloche, Argentina, and Universidad Tecnol\'{o}gica Nacional (UTN-FRBA), Fanny Newbery 111, (8400) Bariloche, Argentina}

\author{Robert M. Turner}
\affiliation{School of Physics and Astronomy, University of Nottingham, Nottingham, NG7 2RD, United Kingdom}

\author{Juan P. Garrahan}
\affiliation{School of Physics and Astronomy, University of Nottingham, Nottingham, NG7 2RD, United Kingdom}

\date{\today}

\begin{abstract}
The dynamics of stochastic systems, both classical and quantum, can be studied by analysing the statistical properties of dynamical trajectories.  The properties of ensembles of such trajectories for long, but fixed, times are described by large-deviation (LD) rate functions.  These LD functions play the role of dynamical free-energies: they are cumulant generating functions for time-integrated observables, and their analytic structure encodes {\em dynamical} phase behaviour.  This ``thermodynamics of trajectories'' approach is to trajectories and dynamics what the equilibrium ensemble method of statistical mechanics is to configurations and statics.  Here we show that, just like in the static case, there is a variety of alternative ensembles of trajectories, each defined by their global constraints, with that of trajectories of fixed total time being just one of these.  We show that an ensemble of trajectories where some time-extensive quantity is constant (and large) but where total observation time fluctuates, is equivalent to the fixed-time ensemble, and the LD functions that describe one ensemble can be obtained from those that describe the other.  We discuss how the equivalence between generalised ensembles can be exploited in path sampling schemes for generating rare dynamical trajectories.
\end{abstract}

\pacs{05.70.Ln, 05.30.Rt, 75.10.Hk}

\maketitle
\section{Introduction}

There is currently a growing interest in the dynamical large-deviations (LD) 
\cite{Eckmann1985,Demboo1998,Ruelle2004,Touchette2009} of many-body systems, both classical \cite{Lebowitz1999,Merolle2005,Giardina2006,*Baule2008,*Gorissen2009,*Jack2010,*Giardina2011,*Hurtado2011,*Nemoto2011,*Bodineau2012,Lecomte2007,Garrahan2007,*Garrahan2009,Hedges2009}
and quantum \cite{Esposito2009,Garrahan2010,Budini2011,*Li2011,*Catana2012,*Gambassi2012,*Flindt2013,*Ivanov2013}.  This is motivated largely by the fact that the emergent, collective dynamics of many-body systems is, in many cases, richer than what can be directly gleaned from their stationary or thermodynamic properties.  That is, dynamics is often more than statics.  In order to fully understand dynamical behaviour one necessarily has to focus on trajectories, and not just on states or configurations. 

A possible method to achieve this is the one we term {\em thermodynamics of trajectories} \cite{Merolle2005,Garrahan2010}.  This is essentially Ruelle's thermodynamic formalism for dynamical systems \cite{Ruelle2004}, adapted to stochastic systems \cite{Lecomte2007}.  It amounts to an ensemble method for trajectories of the dynamics, analogous to the thermodynamic ensemble method for configurations or microstates of standard equilibrium statistical mechanics \cite{Chandler1987,Peliti2011}.  Within this approach appropriate order parameters are time-extensive observables whose fluctuation behaviour characterises the dynamics of the system.  For dynamics, the large-size---or thermodynamic---limit also includes that of large observation time.  This is the regime of large-deviations \cite{Touchette2009}, where the probability distributions of dynamical order parameters (or their generating functions) are fully captured by (large-deviation) functions which play the role of dynamical entropies or free-energies \cite{Touchette2009,Merolle2005,Lecomte2007,Garrahan2010}.  
In particular, the LD functions that correspond to scaled cumulant generating functions of dynamical observables, just like free-energies in the static case, encode in their analytic structure {\em dynamical phase} behaviour.  This has allowed, for example, to reveal phase transitions in the space of trajectories between (equilibrium and non-equilibrium) dynamical phases in systems such as classical glasses \cite{Garrahan2007,Hedges2009}, and other interesting dynamical fluctuation phenomena in a variety of classical and quantum systems \cite{Giardina2006,Garrahan2010,Budini2011}.

One advantage of having an ensemble method for trajectories is that the conceptual and technical machinery developed for equilibrium statistical mechanics \cite{Chandler1987,Peliti2011} can be brought to bear, with appropriate adjustments, on the study of dynamics.  Thinking thermodynamically one question arises very naturally, that of the existence and convenience of alternative trajectory ensembles.  This is the question we address in this paper.  

What is considered most often is a trajectory ensemble with fixed observation time, i.e., one defined by the set of trajectories generated by the dynamics of a certain (and fixed) time duration $\tobs$.  Within such an ensemble it is natural to consider the behaviour of a trajectory observable $K$.  In particular, by controlling the field $s$ conjugate to $K$ (as reviewed below) one can define a ``biased'' ensemble of trajectories---the so-called $s$-ensemble \cite{Hedges2009}---corresponding to the set of actual dynamical trajectories of time extent $\tau$ with an average value of $K$ determined by $s$.  The $s$-ensemble is thus the trajectory ensemble defined by controlling observation time $\tau$ and field $s$.  One alternative ensemble is that of fixed $\tau$ and fixed $K$, and the equivalence between this ensemble and the $s$-ensemble was discussed recently in Ref.\ \cite{Chetrite2013}.  In this paper we consider a different class of trajectory ensembles, those defined not by controlling observation time, but some other trajectory observable such as $K$.  These are {\em fluctuating observation time} ensembles.  We describe how these new ensembles, in their appropriate large ``size'' limit, are also described by corresponding LD functions, and prove their equivalence with fixed $\tau$ ensembles.   Furthermore, these new ensembles allow for potentially significant improvements in numerical simulations of rare trajectories.  

The paper is organised as follows.  In section II we review the $s$-ensemble approach for classical stochastic systems.   In Sect.\ III we introduce the new $x$-ensembles of trajectories with fluctuating time, and in Sect.\ IV we prove the equivalence in the long time limit between the $s$-ensemble and the $x$-ensemble.  Section V extends the $x$-ensemble to the case where multiple dynamical observables are considered, and in Sect.\ VI we generalise our results to the case of open quantum systems.  Section VII describes how to implement the $x$-ensemble numerically with transition path sampling in classical or quantum systems described by continuous time Markov chains, illustrating this method with several examples.   We end in Sect.\ VIII with a general discussion of potential benefits of the $x$-ensemble for path sampling of rare trajectories more generally.

\section{Thermodynamics of trajectories and $s$-ensemble}

Consider a classical stochastic system described by the Master Equation~\cite{Gardiner2004,Peliti2011}
\begin{equation}
\partial_{t} |P(t)\rangle = {\mathbb W} |P(t) \rangle .
\label{ME}
\end{equation}
Here the vector $|P(t)\rangle$ represents the probability distribution at time $t$, 
\begin{equation}
|P(t)\rangle \equiv \sum_{\C} P(\C;t) | \C \rangle
\label{P}
\end{equation}
where $P(\C;t)$ indicates the probability of the system being in configuration $\C$ at time $t$, and $\{ | \C \rangle \}$ is an orthonormal configuration basis, $\langle \C | \C' \rangle = \delta_{\C,\C'}$. For concreteness we focus on continuous time Markov chains, but generalisations of what we describe below are straightforward.
The master operator ${\mathbb W}$ is the matrix
\begin{equation}
{\mathbb W} \equiv \sum_{\C' \neq \C} W({\C \to \C'}) |\C' \rangle \langle \C|
- \sum_{\C} R({\C}) |\C \rangle \langle \C| ,
\label{W}
\end{equation}
where $W({\C \to \C'})$ is the transition rate from $\C$ to $\C'$, and $R({\C})$ the escape rate from $\C$.  In this description, the expectation value of an operator $A$ is given by $\langle A(t) \rangle = \langle - | A | P(t) \rangle$, where $\langle - | \equiv \sum_{\C} \langle \C|$ (such that $\langle - | P(t) \rangle=1$ due to probability conservation).

The dynamics described by Eqs.\ (\ref{ME}-\ref{W}) is realised by stochastic trajectories.  A trajectory of total time $\tobs$ is a time record of configurations, and of waiting times for jumps between them, observed up to a time $\tobs$. That is, if we denote by ${\bf X}_{\tobs}$ such a trajectory, then ${\bf X}_{\tobs} = ( \C_{0} \to \C_{t_1} \to \ldots \to \C_{t_n} )$, where $\C_{0}$ is the initial configuration and $t_{i}$ the time when the transition from $\C_{t_{i-1}}$ to $\C_{t_{i}}$ occurs (so that the waiting time for the $i$-th jump is $t_{i}-t_{i-1}$).  The trajectory ${\bf X}_{\tobs}$ has a total of $n$ configuration changes (and $t_{n} \leq \tobs$, i.e., between $t_{n}$ and $\tobs$ no jump occurred).  Eqs.\ (\ref{ME}-\ref{W}) imply that the probability $P({\bf X}_{\tobs})$ to observe this trajectory out of all the possible ones of total time $\tobs$ is given by
\begin{eqnarray}
P({\bf X}_{\tobs}) &=& 
p_0(\C_{0}) 
\prod_{i=1}^{n}
e^{- (t_{i}-t_{i-1}) R({\C_{t_{i-1}}})}
W({\C_{t_{i-1}} \to \C_{t_{i}}})
\nonumber \\
&& \;\;\;\;\;\;\;\;\;\;\;\;\;
\times e^{- (\tobs-t_{n}) R({\C_{t_{n}}})}
\label{PX}
\end{eqnarray}
with $t_{0} = 0$.  The last factor is the survival probability of the configuration ${\C_{t_{n}}}$ between $t_{n}$ and $\tobs$, and we have also included the probability $p_0$ of the initial configuration.

The properties of the dynamics can be studied by considering the statistics of time-extensive observables \cite{Ruelle2004,Merolle2005,Lecomte2007}.  One such trajectory observable is the ``dynamical activity'', defined as the total number of configuration changes in a trajectory \cite{Lecomte2007,Garrahan2007,Baiesi2009}.  Its distribution over all trajectories ${\bf X}_{\tobs}$ of total time $\tobs$ is
\begin{equation}
P_{\tobs}(K) = \sum_{{\bf X}_{\tobs}} \delta \left( K - \hat{K}[{\bf X}_{\tobs}] \right) P({\bf X}_{\tobs}) 
\label{PK}
\end{equation}
where the operator $\hat{K}$ counts the number of jumps in a trajectory.  For large $\tobs$ this probability acquires a LD form~\cite{Touchette2009,Lecomte2007}, 
\begin{equation}
P_{\tobs}(K) \sim e^{- \tobs \varphi(K/\tobs)}
\label{PKLD}
\end{equation}
Equivalent information is contained in the generating function, 
\begin{equation}
Z_{\tobs}(s) \equiv \sum_{K} e^{- s K} P_{\tobs}(K) 
= \sum_{{\bf X}_{\tobs}} e^{- s \hat{K}[{\bf X}_{\tobs}]} P({\bf X}_{\tobs}) ,
\label{Zs}
\end{equation}
whose derivatives give the moments of the activity, 
$\langle K^{n} \rangle = (-)^{n} \partial_{s}^{n} Z_{\tobs}(s) |_{s=0}$.   For large $\tobs$ the generating function also acquires a LD form~\cite{Touchette2009,Lecomte2007},
\begin{equation}
Z_{\tobs}(s) \sim e^{\tobs \theta(s)} .
\label{ZsLD}
\end{equation}

The analogy with equilibrium statistical mechanics is now evident from Eqs.\ (\ref{PKLD},\ref{ZsLD}).  For the dynamics, the equivalent objects to configurations, or microstates, are trajectories.  Order parameters are time-extensive observables, in this case the activity $K$.  The large-size limit becomes that of large observation time $\tobs$ (more specifically the limit of large space-time volume for a many-body system), and in this limit the order parameter distribution $P_{\tobs}(K)$ is described by the function $\varphi(k)$, which plays the role of, say, a Helmholtz free-energy, which for constant ``volume'' $\tobs$ is only a function of the intensive ``density'' (of the number of transitions) $k = K/\tobs$.  Similarly, $Z_{\tobs}(s)$ is like a partition sum with an associated free-energy $\theta(s)$ (which in this analogy would be like a grand-potential) dependent on the counting field $s$ (akin to a chemical potential for the activity).  Just like thermodynamic potentials, the LD functions $\varphi(k)$ and $\theta(s)$ are related by a
Legendre-Fenchel transformation~\cite{Touchette2009,Lecomte2007}
\begin{equation}
\varphi(k)=-\min_{s}[\theta(s)+ks], 
\label{phi}
\end{equation}
together with the inversion formula
\begin{equation}
\theta(s)=-\min_{k}[\varphi(k)+ks].  
\label{theta}
\end{equation}

The LD function $\theta(s)$ is the quantity of interest.  It is the scaled cumulant generating function for the activity, i.e., the $n$-th cumulant of the activity (per unit time) is given by 
\begin{equation}
\frac{\llangle K^{n} \rrangle}{\tobs} = (-)^{n} \left. \frac{\partial^{n}}{\partial s^{n}}  \theta(s) \right|_{s=0} ,
\label{cumulants}
\end{equation}
where $\llangle \cdot \rrangle$ indicates cumulant (mean, variance, etc.). It thus contains the full statistical information about $K$.  Furthermore, just like a free-energy, its analytic properties encode the the phase structure of the dynamics.  In particular, singularities of $\theta(s)$ are indicative of dynamical, or trajectory, phase transitions.

\begin{figure*}[th]
\includegraphics[width=1.9\columnwidth]{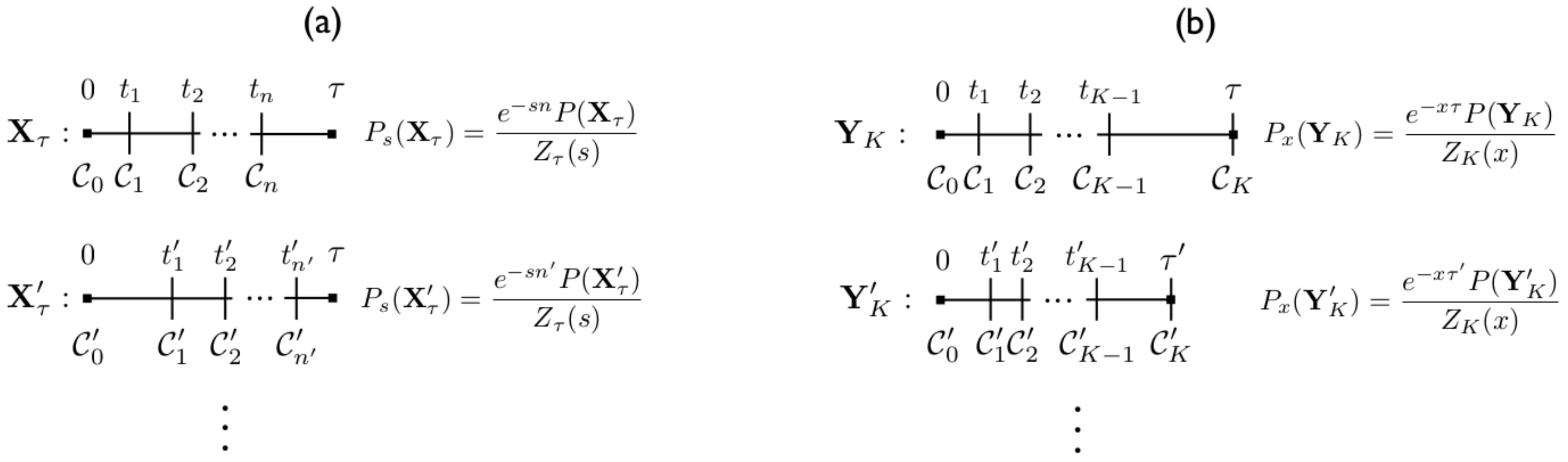}
\caption{(a) The $s$-ensemble is the set of all trajectories that are possible with the dynamics (\ref{ME})-(\ref{W}), of fixed total time $\tobs$, and where the probability of each trajectory is weighed by the number of configuration changes.  It is the ensemble defined by fixed $(\tobs,s)$.  The ensemble of trajectories that corresponds to the actual dynamics (\ref{ME})-(\ref{W}) is given by $(\tobs,0)$.  We sketch two trajectories, the squares indicate the times where trajectories begin and end, and the tick where jumps between configurations take place.  (b)
The $x$-ensemble is the set of all trajectories that are possible with the dynamics (\ref{ME})-(\ref{W}), of fixed number of configuration changes $K$, and where the probability of each trajectory is weighed by the trajectory length.  It is the ensemble defined by fixed $(x,K)$.}
\label{sxensemble}
\end{figure*}

It is useful to clarify the meaning of $s$ at this point.  While $K$ is a physical observable, its conjugate field $s$ is not in principle a physical field (such as pressure or magnetic field); it is a mathematical field which defines the generating function of $K$.  Furthermore, it would appear that only $s=0$ matters.  But to recover all cumulants, derivatives to all orders at $s=0$ are needed, see (\ref{cumulants}), so that the behaviour of $\theta(s)$ for all values $s$ is relevant to the full statistics of $K$.  In the vicinity of $s=0$ the LD function $\theta$ encodes statistical information about all trajectories (as one calculates averages by setting $s=0$) and therefore provides information about {\em typical} dynamics.  In contrast, $\theta$ in the vicinity of $s \neq 0$ carries information about {\em rare} trajectories and thus atypical dynamics.  In particular, we can think of $s$ as defining an ensemble of trajectories whose probability is given by 
\begin{equation}
P_{s}({\bf X}_{\tobs}) \equiv  Z_{\tobs}^{-1}(s)  e^{- s \hat{K}[{\bf X}_{\tobs}]} P({\bf X}_{\tobs}) ,
\label{Ps}
\end{equation}
where ${\bf X}_{\tobs}$ are the same trajectories as the ones generated by the dynamics (\ref{ME}-\ref{W}), but their probability of occurring is now biased by the their activity.  The ensemble given by (\ref{Ps}) is often termed {\em $s$-ensemble}~\cite{Garrahan2007, Lecomte2007, Hedges2009}.  Note that this is an ensemble of trajectories defined by controlling the total time $\tobs$ and the field $s$, so it can be thought of as a $(\tau,s)$-ensemble for trajectories, analogous to a $(V,\mu)$-ensemble for configurations.  
Figure \ref{sxensemble}(a) illustrates this ensemble.

The function $\theta(s)$ can be obtained from a deformation of the master operator $\W$ \cite{Lebowitz1999,Demboo1998,Touchette2009}.  Specifically, for the case of the activity, this deformed operator is \cite{Lecomte2007,Garrahan2007}
\begin{equation}
\W_{s} \equiv \sum_{\C' \neq \C} e^{-s} W({\C \to \C'}) |\C' \rangle \langle \C|
- \sum_{\C} R({\C}) |\C \rangle \langle \C| ,
\label{Ws}
\end{equation}
and $\theta(s)$ is its largest eigenvalue.  For general $s$ the operator $\W_{s}$ does not conserve probability and does not describe a proper stochastic evolution.  It only does so at $s=0$, where it reverts to $\W$, and where $\theta(0)=0$.  The operator $\W_{s}$ is the ``transfer matrix'' for the ``partition sum'' $Z_{\tau}(s)$, that is,
\begin{equation}
Z_{\tobs}(s) = \langle - | e^{\tobs \W_{s}} | p_0 \rangle ,
\label{ZsTM}
\end{equation}
where $| p_0 \rangle$ is the vector for the initial state probability, $| p_0 \rangle \equiv \sum_{\C}  p_{0}(\C) | \C \rangle$.  The above expression is easy to prove from the definitions (\ref{PX}) and (\ref{Zs}). Just like in equilibrium statistical mechanics, this provides a simplification, as calculating $\theta(s)$ then becomes then an eigenvalue problem.

\section{Fluctuating observation time: the $x$-ensemble}

Consider now the case where, instead of keeping fixed the total time $\tau$ of trajectories generated by (\ref{ME})-(\ref{W}), what is kept fixed is the total number of configuration changes $K$, i.e.\ the activity, in each trajectory.  That is, if we denote by ${\bf Y}_{K}$ such a trajectory, then ${\bf Y}_{K} = ( \C_{0} \to \C_{t_1} \to \ldots \to \C_{\tobs} )$, where the number of configuration changes is fixed to be $K$, but the time $\tobs$ of the final $K$-th jump fluctuates from trajectory to trajectory.  From (\ref{ME})-(\ref{W}) the probability of ${\bf X}_{K})$ is
\begin{equation}
P({\bf Y}_{K}) =
p_0(\C_{0}) 
\prod_{i=1}^{K}
e^{- (t_{i}-t_{i-1}) R({\C_{t_{i-1}}})}
W({\C_{t_{i-1}} \to \C_{t_{i}}}) ,
\nonumber
\end{equation}
where $t_{0} = 0$ and $t_{K} = \tobs$.

In analogy with the last section, we ask the question: what is the distribution $P_{K}(\tau)$ of total trajectory length $\tau$ for fixed activity $K$. From the definitions above we have,
\begin{eqnarray}
P_{K}(\tau) &=& 
\sum_{{\bf Y}_{\tobs}} \delta \left( \tau - \hat{\tau}[{\bf Y}_{K}] \right) P({\bf Y}_{K}) 
\label{Pt}
\\
&=&
\sum_{\C_{0} \cdots \C_{K}} 
p_0(\C_{0}) 
\prod_{i=1}^{K-1}
\int_{0}^{t_{i+1}} dt_{i}  
e^{- (t_{i}-t_{i-1}) R({\C_{t_{i-1}}})}
\nonumber \\
&& \;\;\;\;\;\;\;\;\;\;\;\;\;\;\;\;\;\;\;\;\;\;\;\;\;\;\;\;\;\;\;\;\;\;
\times W({\C_{t_{i-1}} \to \C_{t_{i}}}) .
\nonumber
\end{eqnarray}
For large $K$ this probability has a LD form, 
\begin{equation}
P_{K}(\tobs) \sim e^{- K \phi(\tobs/K)} ,
\label{PtLD}
\end{equation}
The corresponding moment generating function for $\tau$ is
\begin{eqnarray}
Z_{K}(x) &\equiv& \int_{0}^{\infty} d\tau e^{- x \tau} P_{K}(\tau) 
\nonumber \\
&=& \sum_{{\bf Y}_{K}} e^{- x \hat{\tau}[{\bf Y}_{K}]} P({\bf Y}_{K}) ,
\label{Zx}
\end{eqnarray}
so that $\langle \tau^{n} \rangle = (-)^{n} \partial_{x}^{n} Z_{K}(x) |_{x=0}$.
For large $K$ the generating function also has a LD form,
\begin{equation}
Z_{K}(x) \sim e^{K g(x)} .
\label{ZxLD}
\end{equation}
The definitions (\ref{Pt})-(\ref{ZxLD}) are the analogous to (\ref{PK})-(\ref{ZsLD}) above: all trajectories have fixed activity $K$ (cf.\ $\tobs$ above); the large limit is that of large $K$ (cf.\ large $\tobs$); the LD function $\phi$ determines the probability of $\tobs$ at large $K$ (cf.\ $\varphi$ for $K$ at large $\tobs$); 
$x$ is the conjugate field to $\tobs$ (cf.\ $s$ and $K$), and the LD function $g$ is the cumulant generating function for $\tobs$ at large $K$ (cf.\ $\theta(s)$ for $K$ at large $\tobs$).  As before, the LD functions $\phi$ and $g$ are related by Legendre-Fenchel transforms,
\begin{equation}
\phi(t) = -\min_{x}[g(x)+tx] , 
\;\;\;
g(x) = -\min_{t}[\phi(t)+tx].  
\label{phitgx}
\end{equation}

Equation (\ref{Zx}) is the ``partition sum'' for the ensemble of trajectories with probabilities
\begin{equation}
P_{x}({\bf Y}_{K}) \equiv  Z_{K}^{-1}(x)  e^{- x \hat{\tau}[{\bf Y}_{K}]} P({\bf Y}_K) .
\label{Px}
\end{equation}
If the $s$-ensemble of the previous section, of fixed $(\tau,s)$, is analogous to an equilibrium $(V,\mu)$ ensemble (since $\tau$ plays the role of volume and $s$ of chemical potential for the activity), then this {\em $x$-ensemble}, of fixed $(x,K)$, can be thought of as analogous to an equilibrium $(p, N)$ ensemble, as $x$ is conjugate to the size trajectories $\tau$ (cf.\ $p$ and $V$ for configurations).
The $x$-ensemble is sketched in Fig.\ \ref{sxensemble}(b).

The generating function $Z_{K}(x)$ can also be written in terms of a transfer matrix operator, 
\begin{equation}
Z_{K}(x) = \langle - | \T_{x}^{K} | p_0 \rangle ,
\label{ZKTM}
\end{equation}
where 
\begin{equation}
\T_{x} \equiv \sum_{\C' \neq \C} \frac{W({\C \to \C'})}{x + R({\C})} |\C' \rangle \langle \C| .
\label{Tx}
\end{equation}
This is obtained by noting from (\ref{Zx}) that $Z_{K}(x)$ is the Laplace transform of $P_{K}(\tau)$, and therefore the elements of $\T_{x}$ are the Laplace transforms of the factors in the integrand of (\ref{Pt}).  The LD function $g(x)$ then corresponds to the logarithm of the largest eigenvalue of $\T_{x}$.

\section{Ensemble equivalence}

Both the $s$-ensemble and the $x$-ensemble are different ways to consider the same underlying dynamics generated by Eqs.\ (\ref{ME})-(\ref{W}).  Just like in the configurational equilibrium case, we expect that in the ``thermodynamic limit'' of large $\tau$ and large $K$ the two ensembles will be equivalent (except perhaps at phase transitions \cite{Chetrite2013}), and that the properties of one ensemble will be obtainable from those of the other.  This equivalence can be proved directly from the spectral properties of the operators $\W_{s}$ and $\T_{x}$. 

The matrices $\W_{s}$ and $\T_{x}$ are directly related to each other.  Specifically, from Eqs.\ (\ref{Ws}) and (\ref{Tx}) we have
\begin{equation}
e^{-s} \T_{x} = \W_{s} \cdot \Q_{x} + \I - x \Q_{x} , 
\label{TW}
\end{equation}
where $\I$ is the identity, $\I \equiv \sum_{\C} |\C \rangle \langle \C|$, and $\Q_{x}$ the diagonal matrix,
\begin{equation}
\Q_{x} \equiv \sum_{\C} \frac{1}{x + R(\C)} |\C \rangle \langle \C|
\label{Qx}
\end{equation}
Consider now a left vector $\langle l |$ that is simultaneously an eigenvector of $\W_{s}$ and $\T_{x}$ with eigenvalues $\theta(s)$ and $e^{g(x)}$, respectively.  If we left multiply (\ref{TW}) by this vector we obtain,
\begin{equation}
\left( e^{-s+g(x)} - 1 \right) \langle l | = \left[ \theta(s) - x \right]\langle l | \Q_{x} . 
\label{TWeq}
\end{equation}
We see that our assumption (of $\langle l |$ being an eigenvector of both $\W_{s}$ and $\T_{x}$) can only be satisfied if $g(x)=s$ and $\theta(s)=x$.  That is, given the function $g$, the function $\theta$ is obtained from its inverse, and vice-versa, 
\begin{equation}
\theta(s) = g^{-1}(s), \;\;\; 
g(x) = \theta^{-1}(x) .
\label{thetag}
\end{equation}
Since the LD rate functions are convex the relation between $g$ and $\theta$ is one-to-one (except perhaps at their boundaries, or at phase-transition points \cite{Chetrite2013}).  Equation (\ref{thetag}) is the statement of the equivalence between the $s$-ensemble and the $x$-ensemble at the level of their respective ``free-energies''.

\begin{figure}[th]
\includegraphics[width=.9\columnwidth]{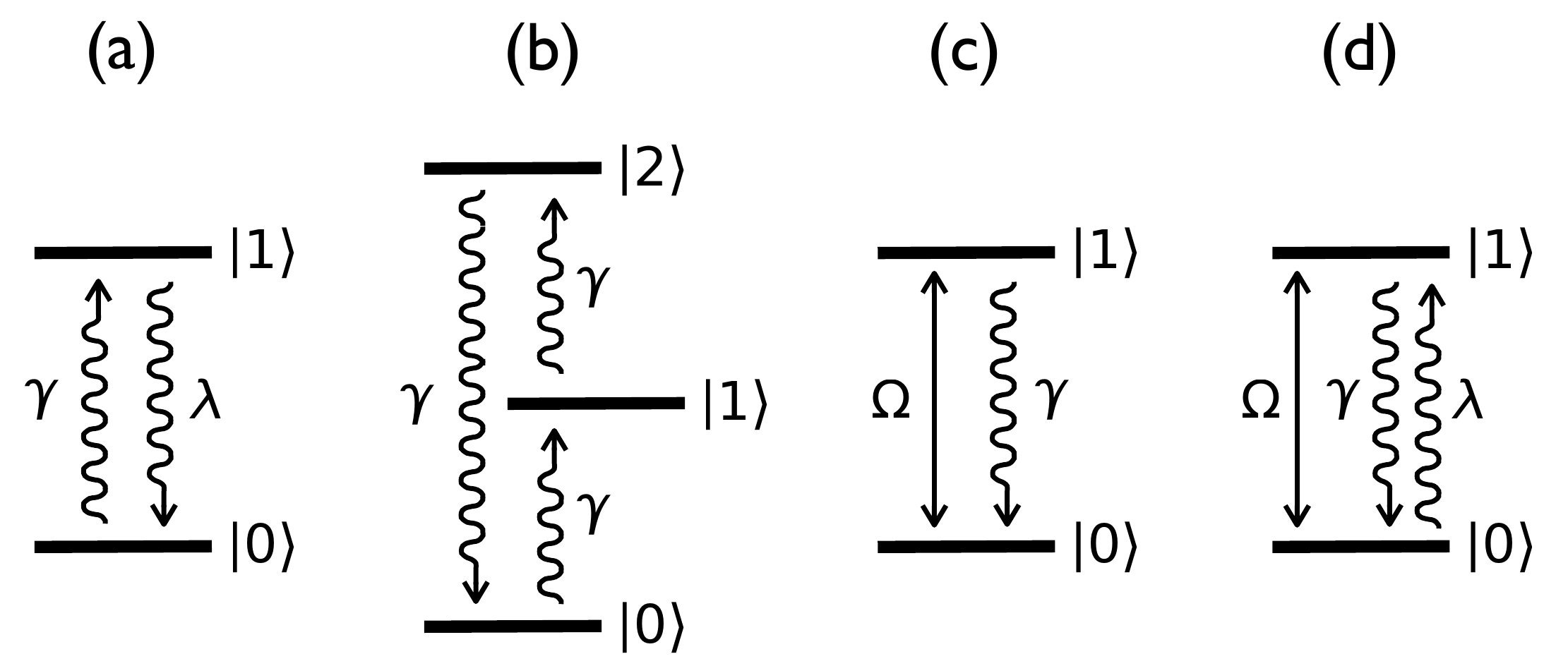}
\caption{(a) Classical two-level system.  (b) Classical three-level system.  (c) Quantum $T=0$ two-level system: the full line indicates a coherent transition of frequency $\Omega$ and the wavy line a dissipative quantum jump of rate $\gamma$ associated to emission into the bath.  (d) Quantum $T 4\neq 0$ two-level system: same as before, but now absorption from the bath leads to a second kind of quantum jump with rate $\lambda$.}
\label{twolevel}
\end{figure}

\subsection{Example: classical two-level system}

As an elementary example consider the classical two level system of Fig.\ \ref{twolevel}(a), where there are only two configurations, $\C \in \{ 0, 1\}$, and the transition rates are, $W({0 \to 1}) = \gamma$ and $W({1 \to 0})=\lambda$.  The operator $\W_{s}$ is,
\begin{equation}
\W_{s} = 
\left(
\begin{array}{cc}
- \lambda & e^{-s} \gamma \\ e^{-s} \lambda & \gamma \\
\end{array}
\right) ,
\end{equation}
and $\theta(s)$ is given by its largest eigenvalue,
\begin{equation}
\theta(s) = 
\frac{1}{2} (\lambda+\gamma) 
\left(
\sqrt{1 - \frac{4 \lambda \gamma}{(\lambda+\gamma)^{2}} (1 - e^{-2s})}
- 1 
\right) .
\label{thetatlc}
\end{equation}
From $\theta(s)$ we can extract the cumulants of the activity.  For the average activity per unit time, i.e.\ the average transition rate between the two levels, we obtain 
\begin{equation}
\frac{\langle K \rangle_{\tobs}}{\tobs} = - \theta'(0) = \frac{2 \lambda \gamma}{\lambda+\gamma} ,
\label{Ktlc}
\end{equation}
as expected.  For the case $\gamma=\lambda$ the LD function reduces to $\theta(s) = \lambda (e^{-s} -1)$, which is the cumulant generating function for a Poisson process with rate $\lambda$.

Similarly, the operator $\T_{x}$ for this problem reads,
\begin{equation}
\T_{x} = \left(
\begin{array}{cc}
0 & \frac{\gamma}{x+\gamma} \\ \frac{\lambda}{x+\lambda} & 0 \\
\end{array}
\right) ,
\end{equation}
and from its largest eigenvalue we obtain the LD function $g(x)$
\begin{equation}
g(x) = \frac{1}{2} \log{\left( \frac{\lambda \gamma}{(x+\lambda)(x+\gamma)} \right)}  .
\end{equation}
This function $g$ is indeed the inverse of the function $\theta$ (\ref{thetatlc}).  The cumulants of the total time are obtained from $g(x)$.  In particular, the average total time, scaled by the number of jumps, is
\begin{equation}
\frac{\langle \tobs \rangle_{K}}{K} = - g'(0) = \frac{\lambda+\gamma}{2 \lambda \gamma} ,
\end{equation}
which is the inverse of (\ref{Ktlc}).  Analogous relations between the moments of $K$ in the fixed $\tobs$ ensemble, and those of $\tobs$ in the fixed $K$ ensemble can be obtained by virtue of Eq.\ (\ref{thetag}).

\section{Generalisation of the $x$-ensemble to multiple observables} 

In the sections above we have proved the equivalence between the $s$-ensemble of fixed observation time $\tau$, where $s$ is conjugate to the overall activity, and the $x$-ensemble of fixed activity $K$, where $x$ is conjugate to the total time.  
This equivalence can be extended to the case where one or more $s$ fields couple to other time-extensive quantities.

Consider a setup as that of Section II, but now we are interested in the statistics of several different time-extensive quantities \cite{Garrahan2007}.  For example, one could think of counting, instead of the total activity, the total number of certain kind of transitions, or the time integral of a certain quantity such as the energy.  Lets say that there are $N$ different dynamical observables, which we denote collectively by the vector $\M \equiv (M_{1}, \ldots, M_{N})$.  Under the dynamics 
Eqs.\ (\ref{ME})-(\ref{W}) there will be a joint probability for observing a combination of these $M$ quantities, $P_{\tobs}(\M)$.  For large $\tau$ this joint probability will have a LD form, 
\begin{equation}
P_{\tobs}(\M) \sim e^{-\tobs \varPhi(\M / \tobs)} ,
\label{PKLDv}
\end{equation}
where the LD function now depends on the whole vector of intensive observables $(M_{1}/\tobs, \ldots, M_{N}/\tobs)$.  The corresponding moment generating function for $\M$ also has a LD form at large $\tobs$ \cite{Touchette2009},
\begin{equation}
Z_{\tobs}(\s) \equiv \sum_{\M} e^{- \s \cdot \M} P_{\tobs}(\M) \sim e^{\tobs \Theta(\s)} ,
\label{ZsLDv}
\end{equation}
where for each observable $M_{n}$ there is a counting field $s_{n}$, collected in the vector $\s \equiv (s_{1}, \ldots, s_{N})$, and where the LD function $\Theta(\s)$ is a now function of this whole vector.  

The partition function $Z_{\tobs}(\s)$ has a transfer matrix representation similar to (\ref{ZsTM}) in terms of an operator $\W_{\s}$.  For simplicity we will assume that the time-extensive observables $\M$ only change at jumps between configurations (extending to cases where observables accumulate in the periods between jumps is straightforward).  In this case $\W_{\s}$ reads \cite{Lecomte2007},
\begin{eqnarray}
\W_{\s} &\equiv& \sum_{\C' \neq \C} e^{- \s \cdot \m({\C \to \C'})} W({\C \to \C'}) |\C' \rangle \langle \C| 
\nonumber \\
&& \;\;\;\;\;\;\;\;\;\;\;\;\;\;
- \sum_{\C} R({\C}) |\C \rangle \langle \C| ,
\label{Wsv}
\end{eqnarray}
where $m_{n}({\C \to \C'})$ is the change in $M_{n}$ under the transition $\C \to \C'$ (for the activity this was just $1$ for all $\C,\C'$ as it counted all transitions equally).  $\Theta(\s)$ is the largest eigenvalue of the operator (\ref{Wsv}).  As before, LD functions are related by Legendre-Fenchel transforms
\begin{equation}
\varPhi(\m)=-\min_{\s}[\Theta(\s)+ \m \cdot \s], 
\;\;\;
\Theta(\s)=-\min_{\m}[\varPhi(\m)+ \m \cdot \s].  
\nonumber
\end{equation}
Eqs.\ (\ref{ZsLDv}),(\ref{Wsv}) define an $(\tobs, \s)$-ensemble for a general set of dynamical order parameters $\M$.

In analogy with Section III, there is an equivalent construct for studying the statistics of $\M$ in trajectories where the total activity $K$ is fixed.  The probability of observing $\M$, together with a total time $\tau$, for a fixed and large $K$ has the form, 
\begin{equation}
P_{K}(\tau,\M) \sim e^{- K \Phi(\tobs/K,\M/K)} .
\label{PtLDv}
\end{equation}
The corresponding moment generating function is
\begin{eqnarray}
Z_{K}(x,\s) &\equiv& \sum_{\M} \int_{0}^{\infty} d\tau e^{- x \tau - \s \cdot \M} P_{K}(\tau,\M) 
\nonumber \\
&\sim& e^{K G(x,\s)} ,
\label{ZxLDv}
\end{eqnarray}
with $\Phi$ and $G$ related by
\begin{eqnarray}
\Phi(t,\m) &=& -\min_{x,\s}[G(x,\s) + t x + \m \cdot \s ] , 
\nonumber \\
G(x,\s) &=& -\min_{t,\m}[\Phi(t,\m)+ t x + \m \cdot \s ].  
\nonumber 
\end{eqnarray}
Equations (\ref{PtLDv}),(\ref{ZxLDv}) define a generalised $x$-ensemble.

The partition sum $Z_{K}(x)$ can be written in terms of a transfer matrix, $Z_{K}(x) = \langle - | \T_{x,\s}^{K} | p_0 \rangle$, with
\begin{equation}
\T_{x,\s} \equiv \sum_{\C' \neq \C} \frac{W_{\s}({\C \to \C'})}{x + R({\C})} |\C' \rangle \langle \C| ,
\label{Txv}
\end{equation}
where $W_{\s}({\C \to \C'})$ are the coefficient of the off-diagonal entries of (\ref{Wsv}).  This allows to prove the ensemble equivalence in this generalised case.  From Eqs.\ (\ref{Qx}),(\ref{Wsv}),(\ref{Txv}) we have that
\begin{equation}
\T_{x,\s} = \W_{\s} \cdot \Q_{x} + \I - x \Q_{x} . 
\label{TWv}
\end{equation}

As in Section IV we search for conditions for which a left vector $\langle l |$ is simultaneously an eigenvector of both $\W_{\s}$ and $\T_{x,\s}$ with eigenvalues $\Theta(\s)$ and $e^{G(x,\s)}$, respectively.  Multiplying $\langle l |$ into = (\ref{TWv}) we get,
\begin{equation}
\left( e^{G(x,\s)} - 1 \right) \langle l | = \left[ \Theta(\s) - x \right]\langle l | \Q_{x} . 
\label{TWeqv}
\end{equation}
It follows that Eq.\ (\ref{TWeqv}) is satisfied when 
\begin{equation}
\Theta(\s) = x_{*}(\s) ,
\label{Thetaxv}
\end{equation}
where $x_{*}(\s)$ is the solution of
\begin{equation}
G(x_{*}(\s),\s) = 0 .
\label{Gsv}
\end{equation}
Equations (\ref{Thetaxv}),(\ref{Gsv}) prove the equivalence between the general $\s$-ensemble and the general $x$-ensemble: they allow to obtain the ``free-energy'' LD functions in one from those in the other, and thus encode the statistical properties of each other.  In the case where $\M$ corresponds only to the activity $K$, as in Sections II-IV, the function $G(x,s) = g(x)-s$, and Eqs.\ (\ref{Thetaxv}),(\ref{Gsv}) reduce to (\ref{thetag}).

\subsection{Example: classical three-level system}

As a  simple example of how the general $s$- and $x$- ensembles relate, consider the classical three-level system in a setup like that of Fig.\ \ref{twolevel}(b).  Suppose we only observe the jumps between configurations $2$ and $0$.  In the notation above we have $N=1$, and $\M$ is just $K_{20}$, the total number of transitions between top and bottom levels.  In the $s$-ensemble, the largest eigenvalue of the operator
\begin{equation}
\W_{s_{20}} = 
\gamma \left(
\begin{array}{ccc}
-1 & 1 & 0 \\ 0 & -1 & 1 \\ e^{-s_{20}} & 0 & -1 \\
\end{array}
\right) ,
\end{equation}
(where $s_{20}$ is the field conjugate to $K_{20}$)
gives the LD function $\Theta(s_{20}) = \gamma (e^{-s/3} - 1)$, which is the cumulant generating function for the number of jumps $K_{20}$ per unit time. 
In the $x$-ensemble context, the relevant operator is 
\begin{equation}
\T_{x,s_{20}} = 
\frac{\gamma}{x+\gamma} \left(
\begin{array}{ccc}
0 & 1 & 0 \\ 0 & 0 & 1 \\ e^{-s_{20}} & 0 & 0 \\
\end{array}
\right) .
\end{equation}
From its largest eigenvalue we obtain the LD function $G(x,s_{20}) = - s_{20}/3 + \log{\gamma} - \log{(x+\gamma)}$.  This is the generating function for cumulants of both $\tobs/K$ and $K_{20}/K$.  If we solve $G(x_{*},s_{20})=0$ for $x_{*}$ we get, $x_{*}(s_{20}) = \Theta(s_{20})$ above, in accordance with (\ref{Thetaxv}),(\ref{Gsv}).

\section{$x$-ensemble in open quantum systems}
\label{Quantumxensemble}

In the previous section we focused for simplicity on stochastic Markovian classical systems.  The extension to Markovian open quantum systems~\cite{Lindblad1976,Plenio1998,Gardiner2004b} is straightforward.  In this case, instead of a master equation for the probability distribution we have a master equation for the density matrix $\rho$
\begin{equation}
\partial_{t} \rho(t) = \WW[\rho(t)] ,
\label{MEq}
\end{equation}
where the {\em quantum master operator} is the super-operator \cite{Gardiner2004b},
\begin{equation}
\WW(\cdot) \equiv -i[H,\cdot] + \sum_{i=1}^{N_{L}} L_i (\cdot) L^\dag_i - \frac{1}{2} \{ L^\dag_i L_i, \cdot \} .
\label{Wq}
\end{equation}
Here $H$ is the Hamiltonian, which generates the coherent part of the evolution, and $L_{i}$ ($i=1,\ldots,N_{L}$) are (bounded) quantum jump operators corresponding to the incoherent effect of the interaction with the environment~\cite{Lindblad1976,Plenio1998,Gardiner2004b}.  The evolution described by (\ref{MEq})-(\ref{Wq}) can be realised by an ``unravelling'' in terms of stochastic wave-functions~\cite{Belavkin1990,Plenio1998,Gardiner2004b}.  This stochastic evolution is given by propagation of the wave-function under the action of the non-Hermitian operator $H_{\rm eff} \equiv H - \frac{1}{2} \sum_i L^\dag_i L_i$, punctuated at random times by ``quantum jumps'' due to the action of the jump operators $L_{i}$.  That is, a quantum version of the continuous time Markov chains discussed above. 

We denote again by $\M \equiv (M_{1}, \ldots, M_{N})$ the time-integrated observables we wish to count, and by $\Theta(\s)$ the large-deviation rate function corresponding to the cumulant generating function for $\M / \tau$ in the large $\tau$ limit.  If under the action of the jump operator $L_{i}$ the observable $M_{n}$ is incremented by $m_{n}^{(i)}$, then the deformed quantum master operator for which $\Theta(\s)$ is its largest eigenvalue reads \cite{Garrahan2010},
\begin{equation}
\WW_{\s}(\cdot) \equiv -i[H,\cdot] + \sum_{i=1}^{N_{L}} e^{- \s \cdot \m^{(i)}} L_i (\cdot) L^\dag_i - \frac{1}{2} \{ L^\dag_i L_i, \cdot \} .
\label{Wsvq}
\end{equation}
This is the open quantum equivalent $s$-ensemble operator to that of Eq.\ (\ref{Wsv}) for the classical case. 

The generalised $x$-ensemble corresponds to controlling the fields $x$, conjugate to the total time $\tau$, and $\s$, conjugate to $\M$, in quantum stochastic trajectories of total and fixed $K$ quantum jumps.  The corresponding LD function $G(x,\s)$ is the largest eigenvalue of the super-operator,
\begin{equation}
\TT_{x,\s}(\cdot) \equiv \sum_{i=1}^{N_{L}} e^{- \s \cdot \m^{(i)}} L_i \left[ ( x + \RR )^{-1} (\cdot) \right] L^\dag_i ,
\label{Txvq}
\end{equation}
where $( x + \RR )^{-1}$ is the inverse super-operator to $( x + \RR )$, i.e., $( x + \RR )^{-1}[( x + \RR )(\cdot)] = (\cdot)$.  Here $\RR$ is the ``escape'' super-operator, $\RR(\cdot) \equiv i H_{\rm eff}(\cdot) - i (\cdot)H_{\rm eff}^{\dagger}$, where $H_{\rm eff}^{\dagger} \equiv H - \frac{1}{2} \sum_{i} L^\dag_i L_i$.  Equation (\ref{Txvq}) is, in the open quantum context, equivalent to Eq.\ (\ref{Txv}) in the classical context.  It is obtained in a similar manner as (\ref{Txv}) by noting that the probability to observe a trajectory of $K$ quantum jumps due to the action of operators $(L_{i_{1}},\ldots,L_{i_{K-1}},L_{i_{K}})$ that occur at times $(t_{1},\ldots,t_{K-1},\tobs)$ is
\begin{eqnarray}
\Tr  \left[ L_{i_{K}} e^{-i (\tobs - t_{K-1}) H_{\rm eff}} L_{i_{K-1}} \cdots L_{i_{1}} e^{-i t_{1} H_{\rm eff}} \rho_{0}
\right.
\nonumber \\
\left.
e^{i t_{1} H_{\rm eff}^{\dagger}} L_{i_{1}}^{\dagger} \cdots L_{i_{K-1}}^{\dagger} e^{i (\tobs - t_{K-1}) H_{\rm eff}^{\dagger}} L_{i_{K}}^{\dagger} 
\right] ,
\end{eqnarray}
where $\rho_{0}$ is the initial density matrix.  

Just like in the classical case, $\WW_{\s}$ and $\TT_{x,\s}$ are directly related, 
\begin{equation}
\TT_{x,\s}(\cdot) = \WW_{\s}[\QQ(\cdot)] + ( \II - x \QQ ) (\cdot) ,
\label{TWvq}
\end{equation}
where $\II$ is the identity super-operator and $\QQ \equiv ( x + \RR )^{-1}$. 
The equivalence between the $s$ and $x$-ensembles is proved in the same manner as before.  We act to the {\em left} on a matrix $\lambda$ which we ask to be simultaneously a left-eigenmatrix of $\TT_{x,\s}$ and $\WW_{\s}$ with eigenvalues $e^{G(x,\s)}$ and $\Theta(\s)$, respectively (where the left action of a super-operator is that of the adjoint), we get
\begin{eqnarray}
(\lambda)\TT_{x,\s} &=& \QQ[ (\lambda)\WW_{\s} ] + ( \II - x \QQ ) (\lambda) 
\nonumber \\
e^{G(x,\s)} \lambda &=& \Theta(\s) \QQ(\lambda)+ \lambda - x \QQ(\lambda) 
\nonumber \\
\Rightarrow
\left( e^{G(x,\s)} - 1 \right) \lambda
&=& 
\left[ \Theta(\s) - x \right] \QQ(\lambda) ,
\label{TWeqvq}
\end{eqnarray}
which has as solutions (\ref{Thetaxv}),(\ref{Gsv}) as in the classical case. 

\subsection{Example: quantum two-level system}
\label{q2level}

As a simple example consider the quantum two-level system of Fig.\ \ref{twolevel}(c), 
corresponding to a system two quantum levels $|0\rangle,|1\rangle$ coherently driven on resonance at Rabi frequency $\Omega$ and coupled to a zero temperature bath~\cite{Gardiner2004}.  The operators that enter in the definition of $\WW$ are the Hamiltonian, 
\begin{equation}
H = \Omega \left( | 0 \rangle \langle 1 | + | 1 \rangle \langle 0 | \right) ,
\end{equation}
and the single jump operator ($N_{L} = 1$)
\begin{equation}
L_1 = \sqrt{\gamma} | 0 \rangle \langle 1 | .
\label{L1}
\end{equation}
We count the number of quantum jumps due to this operator, and consider for simplicity the case where $\gamma = 4 \Omega$ (a particular parameter point where the algebra is simple).  We can write the super-operator $\WW_{s}$ as a matrix \cite{Garrahan2010},
\begin{equation}
\WW_{s} = 
\Omega \left(
\begin{array}{cccc}
 0 & 4 e^{-s}   & i   & -i   \\
 0 & -4   & -i   & i   \\
 i   & -i   & -2   & 0 \\
 -i   & i   & 0 & -2  
\end{array}
\right) ,
\label{Wqtl}
\end{equation}
which acts on the $2 \times 2$ density matrix $\rho$ which we write as the vector,
\begin{equation}
\left(
\begin{array}{c}
\rho_{00} \\ \rho_{11} \\ \rho_{01} \\ \rho_{10}
\end{array}
\right) .
\end{equation}
The largest eigenvalue of (\ref{Wqtl}) is \cite{Garrahan2010},
\begin{equation}
\theta(s) = 2 \Omega (e^{-s/3} - 1) .
\label{Thetaqtl}
\end{equation}
In this matrix form the operator $\TT_{x,s}$ reads,
\begin{equation}
\TT_{x,s} = \frac{4 \Omega e^{-s}}{(x+2 \Omega )^{3}}
\left(
\begin{array}{cccc}
 2 \Omega^{2} & (x+\Omega)^{2}+\Omega^2 & -i x \Omega & i x
   \Omega \\
 0 & 0 & 0 & 0 \\
 0 & 0 & 0 & 0 \\
 0 & 0 & 0 & 0
\end{array}
\right).
\nonumber
\end{equation}
From its largest eigenvalue we obtain, 
\begin{equation}
G(x,s) = - 3 \log{\left(1+ \frac{x}{2 \Omega} \right)} - s , 
\label{Gqtl}
\end{equation}
and it is easy to see that $G[\theta(s),x] = 0$, as expected from Eqs.\ (\ref{Thetaxv}),(\ref{Gsv}).  In particular, $g(x) = G(x,0)$ is the cumulant generating function for the trajectory length $\tau$.  Using the transform (\ref{phitgx}) we obtain the distribution of total time $\tau$ in the large $K$ limit, 
\begin{equation}
P_{K}(\tau) \approx \left( \frac{2 \Omega \tau}{3 K} \right)^{3 K} e^{-2 \Omega \tau + 3 K},
\end{equation}
which is the expected result given that the probability of waiting time between jumps (except for the first one if the initial condition is not $|0\rangle$) is $p(t_{\rm w}) = 4 (\Omega t_{\rm w})^{2} e^{-2 \Omega t_{\rm w}}$.

\section{Path Sampling in the $x$-ensemble}

In this section we describe how to numerically generate the $x$-ensemble, that is, how to obtain ensembles of trajectories weighted by their fluctuating total observation time (and perhaps other observables) by adapting the transition path sampling~\cite{Bolhuis2002} techniques previously developed to study the $s$-ensemble~\cite{Hedges2009}.  We illustrate this approach with a number of examples of classical and quantum stochastic systems ranging from few level systems to many-body glass models, and argue that in most cases the $x$-ensemble based TPS scheme is more efficient than that where the trajectories have a fixed observation time. 

Transition path sampling (TPS)~\cite{Bolhuis2002} is a set of numerical techniques developed to generate so-called reaction trajectories that occur sufficiently infrequently that their direct observation  from simulated dynamics is unfeasible. The rare trajectories description at the heart of the $s$- and $x$-ensemble is therefore well suited to be studied by TPS, and it has had success in numerically describing the $s$-ensemble of a variety of glassy systems~\cite{Hedges2009,Speck2012,Speck2012b}. Rather than biasing the dynamics of a system under study, TPS generates rare trajectories by, in effect, performing a biased random walk in the space of trajectories towards the region of trajectory space compatible with some target trajectory distribution.  

A large part of the TPS literature~\cite{Bolhuis2002} is devoted to techniques for adapting segments of an existing trajectory to propose a new trajectory that is kept or discarded according to the criteria of the biased random walk (usually a metropolis acceptance probability).  It is this process that allows TPS to generate rare trajectories more efficiently than naive sampling of equilibrium dynamics, and great care is taken to ensure it is done without breaking detailed balance with respect to the desired trajectory distribution.  While there are a wide variety of TPS techniques, the most efficient ones tend to be variants of ``forwards-backwards shooting/shifting'', where an existing trajectory is cut somewhere along its history, and one of the two resulting segments is discarded and replaced (perhaps after shifting the remaining segment to the opposite end of the trajectory). It is important to note that replacement of the past of the trajectory requires some form of time-reversed dynamics being run from the cut point.  For stochastic systems in equilibrium, portions of trajectories in the past are easy to generate by simply inverting forwards dynamics.  This is in general not possible out-of-equilibrium, and the inability to achieve time reversal in a simple manner can thus hinder the effectiveness of TPS when a system is driven (which is the typical situation in quantum open dynamics).

\subsection{TPS with fluctuating observation time}

Like in the sections above we consider continuous time Markov chains, both classical or quantum. 
For systems subject classical or quantum master equations, Eqs.\ (\ref{ME})-(\ref{W}) or (\ref{MEq})-(\ref{Wq}), respectively, the standard way to simulate stochastic trajectories is by means of continuous time Monte Carlo~\cite{Landau2009} (often called quantum jump Monte Carlo for the case of open quantum systems~\cite{Plenio1998}).  For the classical case such as scheme amounts to the following~\cite{Landau2009}: (i) given the current configuration of the system $\C$, compute the time $t_{\rm w}$ to the next transition by solving $P_{\C}(t_{\rm w})=r_1$, with $P_{\C}(t_{\rm w}) \equiv e^{- t_{\rm w} R(\C)}$ being the ``survival probability'', $R(\C)$ the escape rate from $\C$, and $r_{1} \in[0,1]$ a uniformly distributed random number; (ii) choose a transition $\C \to \C'$ by drawing a second random number $r_{2}$, where the probability to make the jump $\C \to \C'$ is given by $W(\C \to \C')/R(\C)$; (iii) change the current configuration from $\C$ to $\C'$ and repeat from (i).

In the open quantum case, quantum jump Monte Carlo amounts to an ``unravelling'' of the quantum master equation, Eqs.\ (\ref{MEq})-(\ref{Wq}), that leads to a stochastic evolution for the wave function~\cite{Belavkin1990,Plenio1998,Gardiner2004b}.  Again (quantum) jumps occur stochastically, but in contrast to the classical case the wave function also evolves between jumps through the action of the effective Hamiltonian $H_{\rm eff}$.  That is, if the wave function at $t$ is $\ket{\psi(t)}$, and no quantum jumps occur between $t$ and $t+t_{\rm w}$, then $\ket{\psi(t+t_{\rm w})} = e^{-i t_{\rm w} H_{\rm eff}} \ket{\psi(t)}$.  Furthermore, the survival probability for the waiting time until the next quantum jump is given by    
$P_{\psi}(t_{\rm w}) \equiv ||e^{-i\hbar H_{\rm eff}t_{\rm w}}\ket{\psi(t)}||^2$.  
A stochastic trajectory can then be generated in the following way~\cite{Plenio1998}: (i) given the (normalised) state $\ket{\psi(t)}$, compute the time to the next jump by solving $P_{\psi}(t_{\rm w})=r_1$, where $r_{1} \in[0,1]$ is a uniformly distributed random number; (ii) evolve the wave function by $t_{\rm w}$, $\ket{\psi(t+t_{\rm w})} = e^{-i t_{\rm w} H_{\rm eff}} \ket{\psi(t)}$; (iii) draw a second random number $r_{2}$ to select which quantum jump to perform, where the probability to make the quantum jump $i$ is proportional to $\bra{\psi(t+t_{\rm w})} L_i^\dagger L_i \ket{\psi(t+t_{\rm w})}$;  
(iv) make the selected quantum jump, $\ket{\psi(t+t_{\rm w})} \to L_{i} \ket{\psi(t+t_{\rm w})}$, normalise the resulting state, and repeat from (i). 

The $x$-ensemble is well suited to describe systems whose dynamics is generated by such classical or quantum continuous time Monte Carlo.  A trajectory with a fixed total number $K$ of jumps is fully determined by $K$ pairs of random numbers, $\{r_1,r_2\}^K$.  A TPS scheme for the $x$-ensemble can then be devised based on the method of Ref.\ \cite{Crooks2001}. Since a trajectory is fully encoded in predetermined set of $K$ pairs of random numbers, randomly selecting one of these pairs and modifying it is an efficient way to generate a new trajectory from an old one.  The change in the extensive quantities of interest, $\Delta \tau$ and $\Delta \M$, where $\M$ again denotes the counted observables, can be calculated and the new trajectory accepted or rejected based on the Metropolis acceptance criterion $P_{\rm accept} = \min \{ 1,e^{-(x\Delta \tau + \s \cdot \Delta \M)} \}$.  After a sufficiently large number of accepted moves, a trajectory typified by the fields ($x, \s$) is generated.

Due to the ensemble equivalence demonstrated in Section IV, it is possible to convert $x$-ensemble TPS results, where the values of the fields $(x, \s)$ are fixed, to their equivalent $s$-ensemble described by $\s$ only.  In order to do this it is necessary to find the curve $x^*(\s)$, see Eq.\ (\ref{Gsv}).  This curve passes through the origin of the space spanned by $(x, \s)$, since $G(0,0)=0$ trivially.  It is then possible to move along the $x^*(\s)$ curve by expanding (\ref{Gsv}) for small increments $\delta \s$ in $\s$ which allow to relate the required change in $\delta x$ to the current averages of $\tau$ and $\M$: 
\begin{equation}
\delta x =  \frac{\delta \s \cdot \langle \M \rangle_{x,\s}}{\langle \tau \rangle_{x,\s}} + o(\delta s^2) .
\label{Deltx}
\end{equation}
In order to compute the $s$-ensemble we can therefore start at the point $(x=0, \s=0)$, i.e.\ unbiased dynamics, and progress towards $\s$ by adjusting $x$ according to (\ref{Deltx}), using an $x$-ensemble TPS algorithm for each value of $(x,\s)$.  In this way we recover the properties of ensembles of trajectories with fixed total observation time from simulations of trajectories with fixed number of transitions or jumps.

\subsection{Example: TPS for $T \neq 0$ quantum two-level system}

We again consider the quantum two-level system, but now for the case of $T \neq 0$, Fig.\ \ref{twolevel}(d)~\cite{Plenio1998}.  In most open quantum systems, generating time-reversed dynamics is non-trivial, whether due to the coherent component of the dynamics, or the possibility of an unpaired Lindblad term for which there is no reverse process. In such systems conventional TPS is limited to forward shooting only, limiting its efficiency.  We use the simple case of the $T \neq 0$ two-level problem to illustrate how the $x$-ensemble TPS can efficiently sample such systems.

\begin{figure}
\includegraphics[width=\columnwidth]{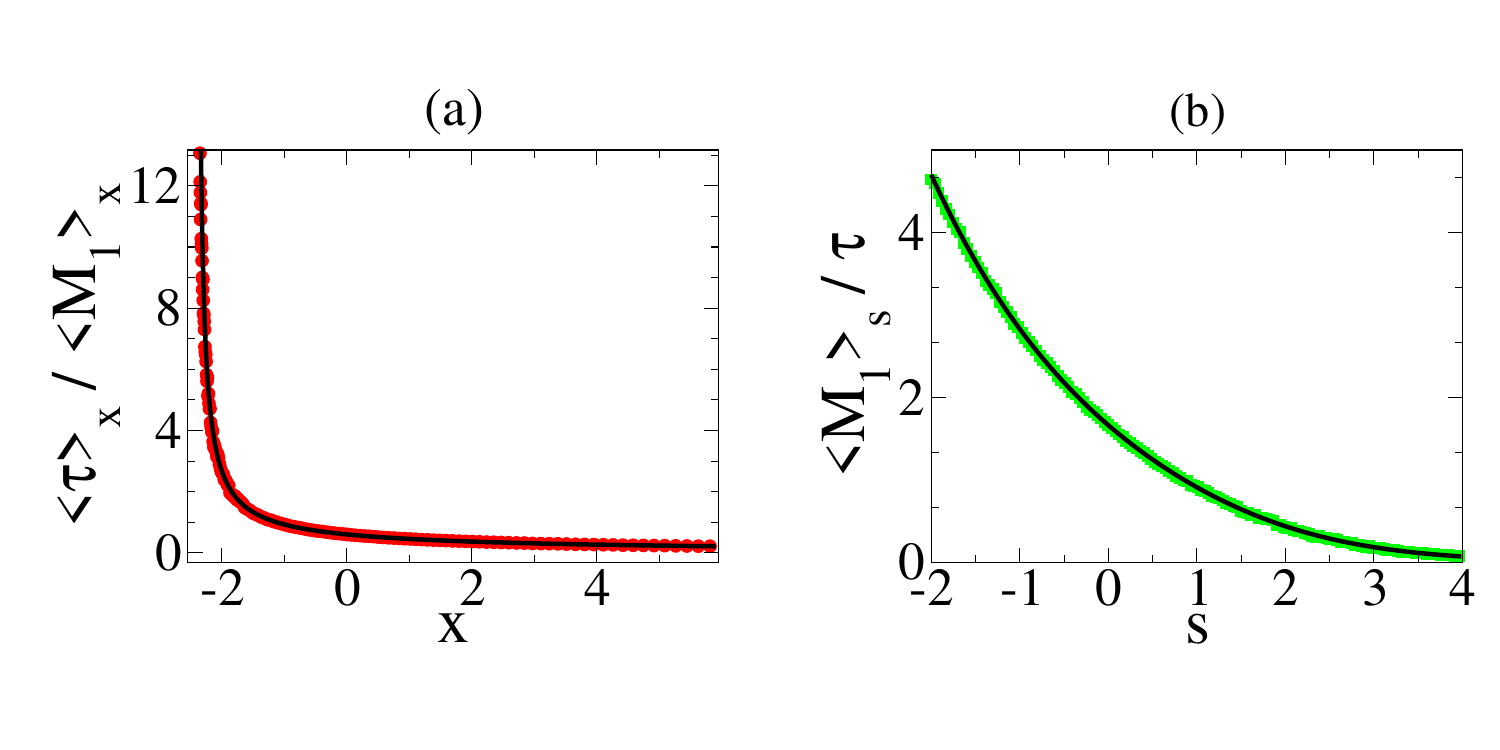}
\caption{$T\neq0$ quantum two-level system. (a) $x$-ensemble: $\langle \tobs \rangle/\langle K_{1} \rangle$ as a function of $x$, along the curve $x^{*}(s_{1})$ in the $x$-ensemble; symbols are from TPS simulations and the curve from exact diagonalisation of $\TT_{x,s_{1}}$.  (b) $s$-ensemble: the symbols are the $s$-ensemble expectation values as a function of $s$, as obtained from the $x$-ensemble TPS simulation, and the curve those from the exact diagonalisation of $\WW_{s_{1}}$.  The parameters here are $\gamma = 6\Omega$ and $\lambda = 2\Omega$.}
\label{fig3}
\end{figure}

Compared to the $T=0$ case of Sect.\ \ref{q2level}, when $T \neq 0$ there is a second jump operator, $L_2 \equiv \sqrt{\lambda}\ket{1}\bra{0}$, 
associated to the absorption of a quanta from the bath, which leads to a projection to the $|1\rangle$ state at a rate $\lambda$ (where the ratio $\lambda/\gamma$ is determined by the temperature $T$).  Lets say we are interested in the statistics of the number of jumps $K_{1}$ due to $L_{1}$ (\ref{L1}).  Using the $x$-ensemble TPS scheme we can compute the average total time, $\langle \tobs \rangle(x,s_{1})$, and the average number of 1-jumps, $\langle K_{1} \rangle(x,s_{1})$, for fixed total jumps $K$ (i.e.\ due to {\em both} $L_{1}$ and $L_{2}$) as a function of the fields $x$ and $s_{1}$.  Figure \ref{fig3}(a) shows the ratio $\langle \tobs \rangle/\langle K_{1} \rangle$ as a function of $x$, along the curve $x^{*}(s_{1})$ (\ref{Thetaxv})-(\ref{Gsv}), comparing the TPS simulation to the exact result.  Figure \ref{fig3}(b) shows the activity associated to $K_{1}$ in the $s$-ensemble, $\langle K_{1} \rangle_{s} = -\theta'(s_1) \tau$, both the exact result from diagonalisation of $\WW_{s_{1}}$, and the numerical estimation from the conversion from the $x$-ensemble TPS simulation, $\langle K_{1} \rangle_{s} = \langle K_{1} \rangle_{x^{*}(s_{1})} \langle \tobs \rangle_{x^{*}(s_{1})}$.

\subsection{Example: micromaser}

Next we consider a micromaser~\cite{Englert2002}, a single-mode cavity coupled to a finite-temperature bath, and pumped by sending excited two-level atoms through the cavity at a constant rate. There are four quantum jump operators associated with the system, two for the atom-cavity interaction  $L_1 = \sqrt{r} \frac{\sin (\lambda \sqrt{aa^\dagger})}{\sqrt{aa^\dagger}}a$, $L_2 = \sqrt{r} \cos (\lambda \sqrt{aa^\dagger})$, and two for the cavity-bath interaction, $L_3 = \sqrt{\kappa}a $, $L_4 = \sqrt{\gamma} a^\dagger $. Here the $a, a^\dagger$ are the raising/lowering operators of the cavity mode, $r$ is the atom beam rate, and $\lambda$ encodes the time of flight of atoms through the cavity. For simplicity, the system can be parameterised by a single ``pump parameter'' $\alpha = \lambda \sqrt{r/(\kappa - \gamma)}$. It can be shown that if the system is initiated in a state with diagonal density matrix, the system stays in a diagonal state. It should be noted that while this effectively reduces the problem to a classical one, generating time-reversed dynamics is still problematic as there is no clear reverse process for action under $L_1$.

The micromaser has a rich trajectory phase diagram with many dynamical phase transitions between states characterised by different photon occupations, $\langle N \rangle$, of the cavity~\cite{Garrahan2011}. 
This complex dynamical phase structure is made manifest by coupling to the number of events $M_1$ under the action of $L_1$, i.e.\ measurements on the atoms leaving the cavity where the atom is in its ground state.  Below we will reproduce this behaviour by converting results from the $x$-ensemble, described by fields $(x,s_1)$, to the equivalent $s$-ensemble described by $s_1$.

Transitions from states with high to low photon occupation occur far more readily than the reverse. 
This can present a problem when trying to numerically recreate a transition away from $x=0$ from a state with low photon occupation to a state with a higher photon occupation. This is similar to the problem of a thermal system becoming stuck in a potential well that is not the global minimum. While there are many techniques to deal with such an issue, such as replica-exchange, there is a more novel approach in the micromaser. Since suitable time-reversed are not available, it is simple to fix the initial state of the trajectory to a large photon occupation, $N$, and set the trajectory length (defined by $K$) large enough that the initial conditions do not have significant impact on the latter stages of the trajectory. Since the micromaser frequently returns to the same state under a TPS algorithm, the $x$-ensemble TPS approach is not punished by large trajectory lengths in the same way the $s$-ensemble is (by returning to a previous state, in effect a constant amount of computation is needed to propose a new trajectory under the $x$-ensemble, where an amount of time of $O(\tau)$ is needed in the $s$-ensemble, since on average half the trajectory is always recomputed; see next section).

\begin{figure}
\includegraphics[width=\columnwidth]{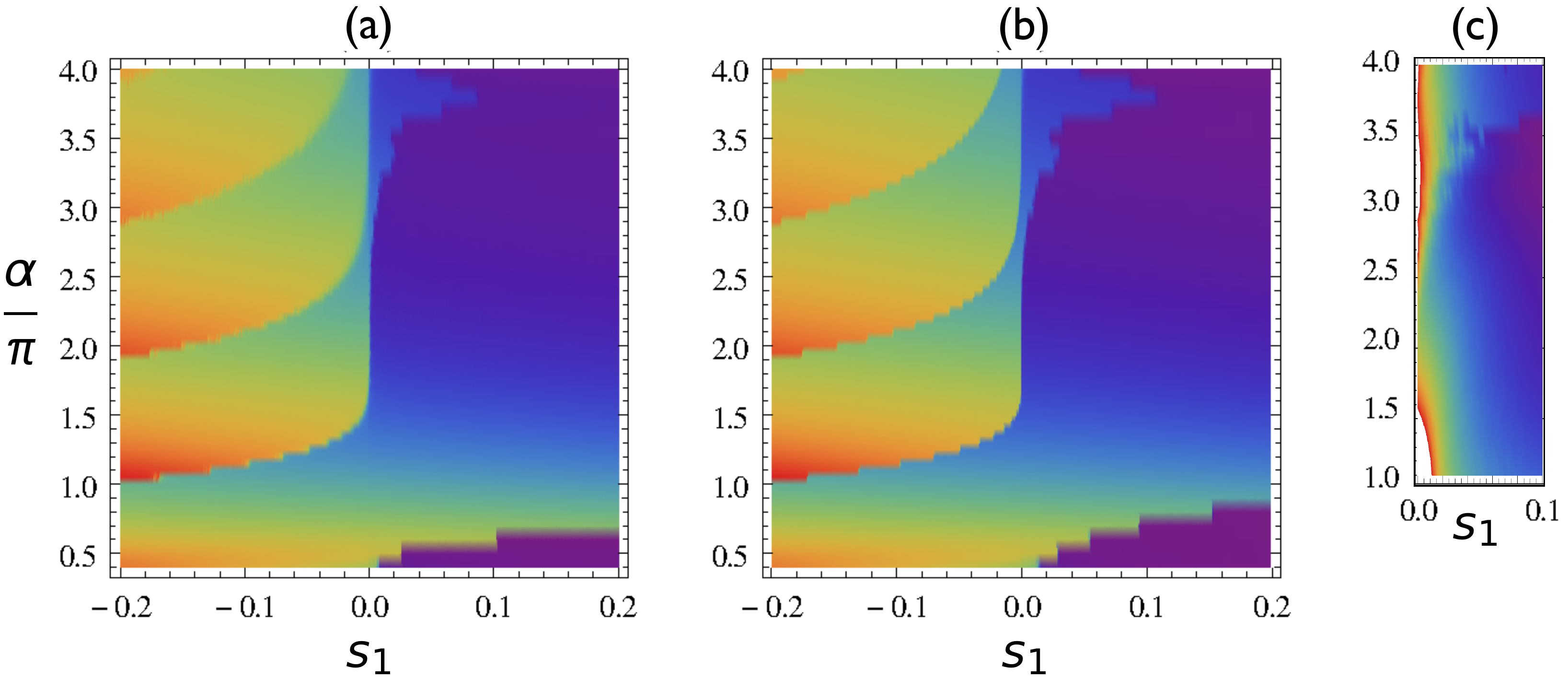}
\caption{Micromaser: average number of $L_{1}$ jumps, $\langle M_{1} \rangle$, as a function of $s_{1}$ and $\alpha/\pi$.  (a) Results from $x$-ensemble TPS of trajectories of fluctuating observation time $\tobs$, transformed to the $s$-ensemble.  (b) Exact numerical diagonalisation of $\WW_{s_{1}}$. Both plots at the same resolution.}
\label{MMxvanalytic}
\end{figure}

Figure \ref{MMxvanalytic}(c) provides a comparison between the efficiency of the $x$-ensemble to that of the $s$-ensemble.  It shows the same quantity as in (a) in the range $s \in [0,0.1]$ and $\alpha \in [\pi , 4 \pi]$ but generated using an $s$-ensemble forward shooting algorithm.  It took about 100 times more computational effort using the $s$-ensemble TPS to generate the data in (c), over a fraction of the parameter range of (a), and clearly the convergence to the exact result is still poor, cf.\ panel (b).  Furthermore, no useful data could be generated in that time using an $s$-ensemble TPS approach for $s_1<0$.  This illustrates the efficiency of the $x$-ensemble TPS scheme as compared to the standard fixed observation time TPS.

\subsection{Example: East facilitated spin model of glasses}

The $s$-ensemble method was first applied \cite{Merolle2005,Garrahan2007} to uncover the dynamical phase structure of kinetically constrained models \cite{Ritort2003}  of glassy systems. Such systems are thermodynamically simple but dynamically complex, and this can be traced back to a singularity in ensemble of trajectories, between ``active'' (equilibrium) and ``inactive'' (non-equilibrium) dynamical phases.  These two phases are stabilised by negative or positive $s$, respectively, with a first-order transition between them at $s=0$ (in the limit of large system size, and for the case where the kinetic constraints are ``hard'', i.e.\ cannot be violated). 

For a demonstration of the functionality of the $x$-ensemble approach in a many-body glassy systems, we consider the East model in one-dimension \cite{Ritort2003}, defined on a lattice of $N$ sites with a binary variable at each site, $n_i = 0,1$ ($i=1,\ldots,N$), and with energy function $E = \sum_i n_i$.  A transition at site $i$, from $0 \rightarrow 1$ with rate $c$, and from $1 \rightarrow 0$ with rate $(1-c)$, can occur only if the neighbouring site to the left is excited, $n_{i-1} = 1$.  This latter condition on the rates is the kinetic constraint.  The transition rates are temperature dependent with $ c = (1+e^{1/T} )^{-1} $. Note that with these definitions detailed balance is obeyed with respect to the Boltzmann equilibrium with energy $E$ at temperature $T$, and since $E$ is non-interacting, despite the strong dynamical interactions, the system evolves towards a non-interacting equilibrium state.   

\begin{figure}
\includegraphics[width=0.9\columnwidth]{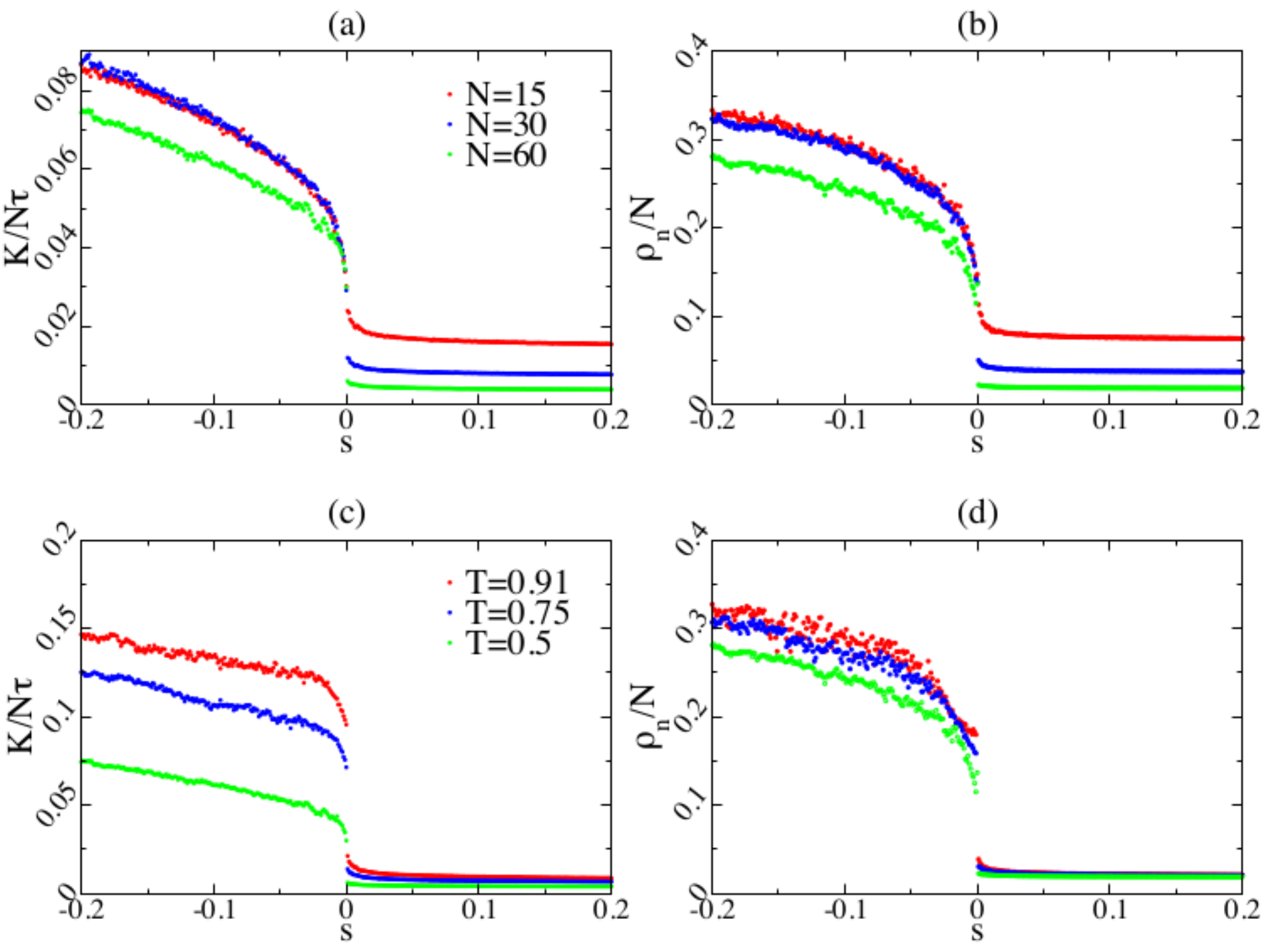}
\caption{(a) The $s$-dependent mean activity per site, $K/N\tau$ and (b) The $s$-dependent mean excitation density per site, $\rho _n/N$.  Both plots are at temperature $T=0.5$, for chain lengths $N=15,30,60$, and the simulations were made using $x$-ensemble TPS. (c/d) Same as before but now for $N=60$ and for temperatures $T=0.91,0.75,0.5$.}
\label{fig5}
\end{figure}

The $x$-ensemble TPS approach is able to efficiently recover the results obtained through $s$-ensemble TPS.  Figure \ref{fig5} shows that we recover the active-inactive crossover, which is seen to get sharper with increasing lattice size, and is present at all temperatures, as expected \cite{Garrahan2007}. Again we have converted from an $x$-ensemble described by field $x$ to an $s$-ensemble described by field $s$.

\subsection{Application of the $x$-ensemble in Transition Path Sampling}

While we have focused on systems simulated with continuous-time Monte Carlo (both classical and quantum), it should be noted that $x$-ensemble TPS is applicable to any system whose trajectories can be fully described by sets of random numbers. The efficiency of the approach will be dependent on the nature of the system under study, but in general there are several factors that should contribute to a greater efficiency in the $x$-ensemble than the equivalent $s$-ensemble TPS schemes.  These factors are the following:

(i) A new trajectory $\Y'$ is generated by altering the $i$-th random number that defines the current trajectory $\Y$ leaving the other random numbers, before and after, unchanged. If at any later stage $j>i$ the trajectory visits the same state as in $\Y$ then no further computation is required to generate $\Y'$.  
This drastically reduces the computation overhead required to propose a new trajectory.  While this recurrence is unlikely in large systems, it will be the main source of efficiency in few-body problems, and perhaps also in systems with large state spaces but with limited dynamical pathways (cf.\ the micromaser). 

(ii) Smoothness of the acceptance criteria. $s$-ensemble TPS has a Metropolis acceptance probability of $P_{\rm accept} = \min{(1, e^{-s\Delta K})}$. Since the number of events is necessarily an integer, a change in activity is only seen if $K$ changes by at least 1. This can lead to low acceptance probabilities, particularly for large $s$. The $x$-ensemble, on the other hand, has a metropolis acceptance probability of  $P_{\rm accept} = \min{(1, e^{-x\Delta \tau})}$ [or more generally for multiple observables, $P_{\rm accept} = \min{(1, e^{-x\Delta \tau - \s \cdot \Delta \M})}$]. Since the trajectory length $\tau$ is continuous small incremental improvements, towards a trajectory typical of the desired value of $(x, \s)$, are more likely to be accepted.

(iii) Trajectories are not altered as drastically. Working in such a manner where trajectories are fully described by a sequence of sets of random numbers, an $s$-ensemble forwards-backwards shooting approach is equivalent to a replacement, on average, of half of the random numbers. While the relationship between the random numbers used to describe a trajectory and the activity of that trajectory is obviously highly non-trivial for a complex system, it is nevertheless to be expected that smaller changes to the random numbers defining the trajectories will have a smaller impact on the activity. By modifying one pair of random numbers only, the $x$-ensemble approach makes smaller incremental improvements facilitating faster convergence.

We expect the $x$-ensemble to be particularly useful in systems that display an active-inactive dynamical phase coexistence, such as glassy systems \cite{Garrahan2007,Hedges2009}.  Using a fixed-time trajectory, as in the $s$-ensemble, the length of the trajectory needs to be set long enough that any interesting behaviour in the inactive phase is captured.  But this can lead to an unnecessarily large amount of information on the active phase being recorded.  In contrast, in the $x$-ensemble little computation time is wasted on the active phase: by fixing total event numbers, the same quality of statistics is generated for both active and inactive dynamics.

\bigskip

\acknowledgments

We are grateful to Phill Geissler for discussions.  AAB thanks the kind hospitality of the School of Physics \& Astronomy during his stay at the University of Nottingham.  This work was supported by CONICET, Argentina, under grant no.\ PIP 11420090100211, and by 
Leverhulme Trust grant no.\ F/00114/BG.


%

\end{document}